\documentclass{jnmp}

\usepackage{graphicx}

\JNMPnumberwithin{equation}{section}

\def\a{\alpha}
\def\b{\beta}

\def\de{\delta}   
\def\eps{\varepsilon}
\def\la{\lambda}

\def\om{\omega}
\def\th{\theta}
\def\vth{\vartheta}
\def\vphi{\varphi}

\def\R{{\bf R}}

\def\<{\langle}
\def\>{\rangle}

\def\({\left(}
\def\){\right)}
\def\[{\left[}
\def\]{\right]}
\def\=#1{\bar #1}
\def\~#1{\widetilde #1}
\def\.#1{\dot #1}
\def\^#1{\widehat #1}
\def\"#1{\ddot #1}

\def\beq{\begin{equation}}
\def\eeq{\end{equation}}

\def\L{{\mathcal L}}

\def\Y{Yakushevich }
\def\EOR{\hfill $\odot$}

\begin{document}

\renewcommand{\evenhead}{M Cadoni, R De Leo and G Gaeta}
\renewcommand{\oddhead}{Solitons in a double pendulum chain model}

\thispagestyle{empty}

\FirstPageHead{*}{*}{2006}{\pageref{firstpage}--\pageref{lastpage}}{Article}

\copyrightnote{2006}{M Cadoni, R De Leo and G Gaeta}

\Name{Solitons in a double pendulums chain model, and DNA
roto-torsional dynamics\footnote{Work supported in part by the
Italian MIUR under the program COFIN2004, as part of the PRIN
project {\it ``Mathematical Models for DNA Dynamics ($M^2 \times
D^2$)''}.}}

\label{firstpage}

\Author{Mariano Cadoni $^{\dag}$, Roberto De Leo $^{\dag}$, and
Giuseppe Gaeta $^{\ddag}$}

\Address{$^\dag$ Dipartimento di Fisica, Universit\`a di Cagliari,
and INFN, Sezione di Cagliari, Cittadella Universitaria, 09042
Monserrato, Italy; \\
{\rm E-mail addresses:} {\tt mariano.cadoni@ca.infn.it}
{\rm and} {\tt roberto.deleo@ca.infn.it} \\
$^\ddag$ Dipartimento di Matematica, Universit\`a di Milano, via
Saldini 50, I--20133 Milano, Italy; {\rm E-mail address:} {\tt
gaeta@mat.unimi.it}}

\Date{Received August 3, 2006; Revised September 29, 2006;
Accepted Month *, 200*}

\begin{abstract}
\noindent It was first suggested by Englander et al to model the
nonlinear dynamics of DNA relevant to the transcription process in
terms of a chain of coupled pendulums. In a related paper
\cite{CDG} we argued for the advantages of an extension of this
approach based on considering a chain of double pendulums with
certain characteristics. Here we study a simplified model of this
kind, focusing on its general features and nonlinear travelling
wave excitations; in particular, we show that some of the degrees
of freedom are actually slaved to others, allowing for an
effective reduction of the relevant equations.
\end{abstract}

\section*{Introduction and motivation}

In a seminal paper appeared a quarter century ago, Englander,
Kallenbach, Heeger, Krumhansl and Litwin \cite{Eng} suggested that
DNA solitons, i.e. nonlinear mechanical excitations of the DNA
double helix, could have a key role in DNA functional processes --
such as DNA transcription and replication -- in that they would
provide automatic focusing of energy and synchronization of
opening along the chain. They also were the first to suggest a
simple mechanical model to illustrate their argument; this
consisted of a one-dimensional chain of simple pendulums, each of
them coupled to the neighboring ones. In this way travelling
solitons were described by the sine-Gordon equation, which is
known to support both topological and dynamical solitons. (Note
that, contrary to the Davydov soliton in alpha-helices \cite{Dav}
which is intrinsically quantum, in this case we deal with
classical mechanics and hence classical solitons.)

Following the suggestion of Englander et al., a number of models
for the nonlinear dynamics of DNA have been elaborated by
different scientists in later years.

Research was pursued essentially in two directions: on the one
hand, in DNA {\it denaturation} the relevant aspect is the
separation between the two helices, and one looks mainly at the
degree of freedom describing distance between the two bases in a
(Watson-Crick) pair. Considerable progress in this direction has
been obtained via the model introduced by Peyrard and Bishop
\cite{PB}; this has then been refined by Dauxois \cite{Dauhel} and
extended in the BCP model by Barbi, Cocco and Peyrard
\cite{BCP,BCPR,CM,PeyHouches} (related models are discussed in
\cite{Joy,Yakhel}). See e.g. \cite{Theo,PeyNLN} for a discussion
of results and recent advances in this direction, and
\cite{CuS,Rap,TPM} for matters related to (thermodynamic)
stability and bubble formation.

On the other hand, models for the roto-torsional dynamics of DNA
have been considered by other authors; the torsional degrees of
freedom are relevant in connection with the opening of the DNA
double helix taking place to allow RNA-Polymerase to access the
base sequence to transcript genetic information \cite{YakuBook}.
Several models have been proposed in this direction; references
and some detail on these can be found in
\cite{GRPD,YakuBook}.\footnote{It should be stressed that these
models deal with the DNA double helix alone, i.e. do not consider
its interaction with the environment or other agents as RNA
Polymerase; as such, they are not of direct relevance for
processes involving other actors, albeit they are definitely a
first step in the study of these (see e.g. \cite{PeyNLN,YakuBook}
for a discussion of the possible functional role of nonlinear
excitations in DNA). On the other hand, the description they
provide of the dynamics of the DNA molecule could nowadays be
tested, in principles, by means of single-molecule experiments
\cite{Lav,Rit}.}

A particularly simple model (Y model) was proposed by Yakushevich
\cite{YakPLA,YakPhD}; see also \cite{Gaehel,GaePLA2,GaeJBP,GRPD}.
Despite its simplicity, the Y-model succeeds in describing several
relevant features of the DNA nonlinear (and linear) dynamics
related to the relevant processes \cite{GRPD,YakuBook,YakPRE}, and
is thus the subject of continuing interest. In recent works it has
been shown that all the (sometimes, very crude) approximations
introduced by \Y in her model -- some of these substantially
affecting the dispersion relations for the model \cite{GML} --
have very little impact on the fully nonlinear dynamics and in
particular on solitons' shape \cite{GY1,GY2}.

The Y model should in any case be seen as only a first step in a
hierarchy of increasingly accurate models \cite{YakPhD}, and has
some drawbacks which should be removed by considering more
detailed versions of the model; these are both quantitative and
conceptual.

On the quantitative side, we mention the impossibility to fit the
observed speed of transversal waves in DNA with physical values of
the parameters \cite{YakPRE}; and the fact that soliton speed
remains -- provided it is smaller than a maximal speed --
essentially a free parameter \cite{GaeSpeed}.

On the conceptual side, solitons are possible in these models
thanks to the homogeneous character of the chain; but we know that
actual DNA is strongly inhomogeneous, as bases are quite different
from each other, and homogeneous DNA would not carry any
information.\footnote{The idea behind considering such a model is
of course that the homogeneous model can be considered as an
``average'' version of a more realistic model, which could be
studied perturbatively; but in a model like the Y model (and more
generally those considered in the literature) a perturbation
breaking translational invariance would destroy the soliton
solutions.}

The model we consider here describes the DNA double chain via two
(rotational) degrees of freedom per nucleotide, hence it will
called a ``composite Y model'', following the nomenclature
introduced in \cite{CDG}. These degrees of freedom are related
separately to rotation of the sugar-phosphate backbone (which can
be of any magnitude) on the one hand, and of the nitrogen bases
(which are constrained to a limited range) on the other hand. Both
rotations are in the plane perpendicular to the double helix axis.

This model is a simplified -- and somehow a ``skeleton'' --
version of the more realistic (and involved) model considered in
\cite{CDG}: in that paper we triggered our model to the actual DNA
geometry and dynamics, while here we consider a simplified model
so to focus on the general abstract mechanism of interaction
between topological and non topological degrees of freedom. For
the same reason we will restrict to motions which are symmetric
under the exchange of the two chains (as often done also in the
analysis of the Peyrard-Bishop and the BCP models).

The model we consider, and more generally the class of composite Y
models \cite{CDG}, represents an improvement with respect to the
standard Y model both from  a quantitative and qualitative point
of view.

On the quantitative, phenomenological, side, we have a much higher
flexibility of the chain described by the model and dramatic
consequences on the model ability to provide realistic physical
quantities. E.g., with the composite Y model considered in
\cite{CDG}, one obtains the experimentally observed transverse
phonon speed using interaction energies of the physical order of
magnitude\footnote{As mentioned above, within the standard Y model
the same speed can be fitted only using an intrapair interaction
energy which is about 6000 times the physical value
\cite{YakPRE}.}, and a selection of solitons' speed.

On the qualitative side, which is maybe even more interesting (and
more widely applicable than merely DNA), the composite Y model is
remarkable in that the uniform and the non-uniform parts of the
DNA molecule (backbone and bases respectively) are described by
separate degrees of freedom. It happens that, as a consequence of
the geometry of the DNA molecule, the degree of freedom describing
the backbone supports topological -- hence strongly stable --
solitons, while the one describing the motion of bases performs
quite limited excursion due to steric hindrances; as mentioned
above, this is taken into account in our model. It is thus quite
conceivable that introducing in the model a non-uniformity which
affects only the latter degree of freedom, a perturbation approach
would allow to obtain solutions in terms of perturbed solutions
for the uniform model\footnote{It should be mentioned that the BCP
model \cite{BCP,BCPR,PeyNLN} also presents the interaction of
topological and non-topological degrees of freedom; however, there
the topological degree of freedom is actually a cyclic variable
and one has correspondingly a conservation law: this leads to a
dynamics less rich topologically.}; see sections \ref{slaved} and
\ref{discussion} below.

The perturbative approach is also attractive in that by a suitable
limiting procedure, see section \ref{perturbative} below (and
\cite{CDG}), the composite Y model reduces to a standard Y model,
for which exact solutions can be obtained. Thus, a perturbative
description for the uniform composite Y model can be obtained by
perturbing the system near the standard Y solutions.

A drawback of general composite Y models is that the equations
describing its dynamics are too complex to be solved analytically,
and even the perturbative expansion around the solitonic Y
solution can be very hard to control \cite{CDG}. On the other
hand, most of the nice features of the composite model seem to be
quite generic, related mostly to doubling of the degrees of
freedom and their different topological features, i.e. largely
independent of the model details. Numerical investigations for the
``realistic'' composite Y model of \cite{CDG} showed that
introducing a number of simplifications into the model does not
affect its main features -- confirming the observations for the
standard Y model \cite{GY1,GY2} -- but surely makes it easier to
handle it at the analytical level.

Motivated by the previous arguments, we study in this paper a
simple -- possibly the simpler -- composite Y model, for which the
comparison with the standard Y model is immediate. Our main
purpose is indeed to focus on the essential features introduced by
having two different kinds of degrees of freedom in chain models.

For the sake of concreteness we make reference to -- and discuss
the consequences for -- DNA dynamics, but it will be quite clear
that most of our discussion is more general.


\section{The model}

Let us now describe our model. We define this in abstract terms,
but it can be useful -- in order to fix ideas -- to recall that in
applying it to DNA the ``first pendulums'' mentioned below and
associated to angles $\th$ represent the nucleosides (segment of
the phosphodiester chain and the attached sugar ring), while the
``second pendulums'' associated to angles $\phi$ represent the
nitrogen bases.

\def\th{\theta}

We consider two chains of double pendulums. On each chain axis
there are equally spaced sites at points of coordinate $z = n
\de$, with $\de$ a dimensional constant and $n \in {\bf Z}$.

At each site $i \in {\bf Z}$ on each chain axis, there are
identical simple\footnote{That is, pendulums consisting of a
perfectly rigid and massless bar rotating about an extremum, and
of a point mass fixed on the opposite extremum.} "first pendulums"
of length $R$ and mass $M$, which can rotate with no limitations
in the plane orthogonal to the axis. The rotation angle of the
first pendulum on chain $a=\pm1$ at site $i$ will be $\th^a_i$.

\begin{figure}
  \includegraphics[width=150pt]{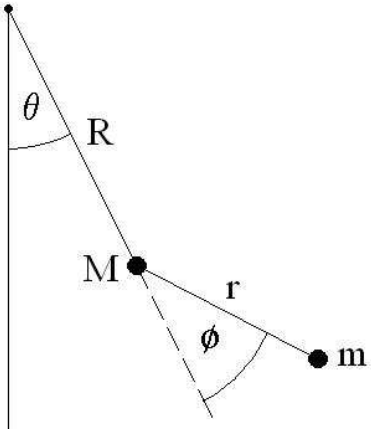}\\
  \caption{Notation for each of the double pendulums along the two chains; see text.}
  \label{dbpend}
\end{figure}

Attached to the point mass of each "first pendulum" there is a
simple "second pendulum" of length $r$ and mass $m$, which can
rotate in the same plane as the first pendulums. The angle of
rotation of the second pendulum (with respect to the direction of
the first pendulum) will be $\phi^a_i$. Details are shown
pictorially in fig.\ref{dbpend}.

The second pendulum is {\it not} free to swingle through a full
circle, but is instead constrained to stay in the range $|\phi^a_i
| \le \phi_0$. This constraint will also be modelled by adding a
constraining potential, i.e. a potential $V_c (\phi)$ which has
the effect of limiting {\it de facto} the excursion of the $\phi$
angles.

The reason for this limitation in the $\phi$ range is our
intention to model the DNA molecule: in that the bases cannot move
outside a certain range, as they would otherwise collide with
other parts of the molecule (the phosphodiester chain or the sugar
ring).

As for the relative position of the two double pendulums chains,
this is such that the distance between the suspension points at
the same site -- and thus with the same $z$ coordinate -- is $2
A$. Equivalently, at rest (i.e. for $\th^{(a)}_i = \phi^{(a)}_i =
0$ for all $a=1,2$ and $i \in {\bf Z}$), the distance between
masses at the end of second pendulums at the same site on the two
chains is $\ell_0 := 2 (A - R - r)$.

Let us denote, dropping for a moment the index $i$ and writing $a
= \pm$ instead of $a=1,2$, as $(x,y)$ the cartesian coordinates --
in the plane orthogonal to the double helix axis $z$, with origin
in the central point -- of the the point mass at the end of the
second pendulum.

In the equilibrium position, given by $\th=\phi=0$, we have
$(x,y)=(x_0,y_0)$ with $x^\pm_0 = \mp A \pm R \pm r$, $y_0 =0$. In
general, \beq\label{positions} x^\pm  = \mp A \pm R \cos \th \pm r
\cos (\th + \phi ) \ , \ \ y^\pm = \pm R \sin \th \pm r \sin (\th
+ \phi ) \ . \eeq

\medskip\noindent
{\bf Remark 1.} In the standard \Y model, one considers the
approximation $\ell_0 \to 0$; it is known that this approximation
has a strong impact on linearized dynamics and dispersion
relations \cite{GML}, but for what concerns fully nonlinear
dynamics it has been observed that -- at least in the framework of
the \Y model -- the approximation has a very little impact on
travelling solitons \cite{GY1}. This suggests the possibility of
considering the same approximation $\ell_0=0$, hence $A=R+r$, here
as well. We will indeed adopt this. \EOR

\subsection{The Lagrangian}

Let us now describe the different terms appearing in the
Lagrangian $\L$ for the two chains of double pendulums.

The kinetic energy of the double pendulum (i.e. a disk and the
attached pendulum) at site $i$ on chain $a$ is $ T^a_i = (1/2) [ I
({\dot \th}^a_i)^2 + m (({\dot x}^a_i)^2 + ({\dot y}^a_i)^2) ]$;
using (\ref{positions}) to express $x^a_i$ and $y^a_i$ in terms of
$\th = \th^a_i$ and $\phi = \phi^a_i$, we get with simple algebra
\beq\label{kin} T^a_i \ = \ (I/2) {\dot \th}^2 \, + \, (m/2)
\left[ R^2 {\dot \th}^2 + 2 R r \cos \phi \( ({\dot \th})^2 +
{\dot \th} {\dot \phi} \) + r^2 ({\dot \th} + {\dot \phi})^2
\right] \ .  \eeq The total kinetic energy is of course $T =
\sum_{a,i} T^a_i$.

We now pass to consider coupling between different double
pendulums; we will have only nearest-neighbor interactions, which
will be of three different types\footnote{In analyzing small
amplitude dynamics, the ``helicoidal'' coupling between disks $h$
sites away from each other (with $2 h \de $ the pitch of the DNA
helix) would also play a role \cite{Dauhel,Gaehel,GaeJBP,GRPD}.
Physically this interaction is mediated by Bernal-Fowler filaments
\cite{Dav}, and it can be described by a potential $$ U_h^{(i)} \
= \ (1/2) \, K_h \, [ (\th^{(+)}_{i+h} - \th^{(-)}_i )^2 \, + \,
(\th^{(-)}_{i+h} - \th^{(+)}_i )^2 ] \ ; $$ however, this is
unessential in the fully nonlinear, i.e. large amplitude, regime
and will hence not be considered here, as we focus on fully
nonlinear excitations (in the small amplitude regime, introducing
this term removes a degeneration of the system and is therefore
qualitatively relevant).}.

\medskip\noindent {\tt (1)} A torsional coupling between successive
disks on each chain, described by a potential \beq\label{pot}
U_t^{(a,i)} \ = \ K_t \, [ 1 - \cos (\th^a_{i+1} - \th^a_i ) ] \ ;
\eeq the total torsional potential is of course $V_t = \sum_{a,i}
U_t^{(a,i)}$. This, albeit entirely natural in terms of the
mechanical model (and thus mentioned here), is actually of lesser
relevance in modelling DNA and will thus be overlooked; see Remark
3 below.

\medskip\noindent {\tt (2)} A stacking coupling between successive
pendulums on each chain, described by a potential $ U_s^{(a,i)} =
(1/2) K_s [ (x^a_{i+1} - x^a_i)^2 + (y^a_{i+1}-y^a_i)^2 ]$. This
is expressed in terms of the $\th^a_i$ and $\phi^a_i$ angles,
using again (\ref{positions}) and with $\la = r/R$, as
\beq\label{stack}
\begin{array}{rl} U_s^{(a,i)}
 =&  ( K_s R^2 / 2) \, \[ \( \cos \th^a_i - \cos \th^a_{i+1} + \la [ \cos
(\phi^a_i + \th^a_i )  - \cos( \phi^a_{i+1} + \th^a_{i+1})] \)^2
\right. \\ & \left. + \( \sin \th^a_i - \sin \th^a_{i+1} + \la [
\sin( \phi^a_i + \th^a_i)  - \sin( \phi^a_{i+1} + \th^a_{i+1} )]
\)^2 \] \ ;
\end{array} \eeq the total stacking potential is of course $V_s =
\sum_{a,i} V_s^{(a,i)}$.

\medskip\noindent {\tt (3)} A pairing coupling between double
pendulums at the same site on opposite chains. We assume this is
due to an interaction between the point particles at the extremal
points of the pendulums, described by a potential \beq\label{pair}
U_p^{(i)} \ = \ F ( \rho_i ) \eeq where $\rho_i$ is the distance
between point masses of the two pendulums at site $i$ on the two
chains, $ \rho_i := [ (x_i^{(+)} - x_i^{(-)} )^2 + (y_i^{(+)} -
y_i^{(-)} )^2 ]^{1/2} $. The total pairing potential will of
course be $V_p = \sum_{i} U_p^{(i)}$.

Recent work of ours in the framework of the simple \Y model
\cite{GY1,GY2} has shown that the shape of travelling topological
solitons is very little affected by the detailed shape of the
intrapair potential \cite{GY2}. We will thus adopt the simplest
choice, i.e. an harmonic potential $ U_p^{(i)} = (1/2) K_p
\rho_i^2 $; this becomes anharmonic when expressed in terms of the
$\th^a_i$, $\phi^a_i$ rotation angles, \beq\label{pairlinear}
U_p^{(i)} = 4 K_p \[ (R+r)^2 -  r R (1 - \cos \phi^a_i )
 - (r+R) \( r \cos (\phi^a_i + \th^a_i ) + R \cos \th^a_i
\) \] \ . \eeq

\medskip

Putting all these together, and considering also the constraining
potential $V_c$ to be discussed below, the Lagrangian describing
our system will therefore be \beq\label{lagr} \L \ = \ T \ - \ \(
V_t + V_s + V_p + V_c \) \ . \eeq

Some remarks are in order concerning the potential interactions
introduced above.

\medskip\noindent
{\bf Remark 2.} We have explicitly assumed harmonic potentials for
the torsional and stacking coupling between successive elements
(disks and pendulums respectively) on the same chain. It should be
mentioned, in this context, that recent work by Saccomandi and
Sgura has shown that the choice of a harmonic potential for the
interpair interactions (torsional and stacking couplings) is not
without effect: considering non-harmonic couplings leads to a
stronger energy localization \cite{SaSg}. We will defer
consideration of the effect of anharmonic interpair couplings in
our model to later work. \EOR

\medskip\noindent
{\bf Remark 3.} We also note that we have considered here both a
torsional coupling originating in the coupling between adjacent
first pendulums (in DNA, in the phosphodiester chain) and a
rotational interaction originating in interaction between adjacent
second pendulums (in DNA this models the stacking interaction
between neighboring bases).

The physical origin of these in actual DNA molecule is rather
diverse. Indeed, the stacking interaction is essentially due to
polar interaction between the bases; these are essentially flat
and rigid assemblies of atoms with a strong dipolar momentum, and
correspondingly we have (in physico-chemical notation) a $\pi-\pi$
bond. On the other hand, the segment of a hypothetical
phosphodiester chain without attached bases would be essentially
free to rotate with respect to each other (if not for rather small
polar forces), the main obstacle to such motion originating in
hindrances caused by bases and their occupied volume and polar
moments.

It is thus reasonable, as announced above (and also in view of the
considerable complications of equations that would be obtained
from the full Lagrangian and the fact the two motions are strongly
related geometrically), to consider the approximation $ K_t \to 0
$, as we do from now on. This corresponds to the well known fact
that the forces stabilizing the DNA double chain are essentially
the interbase pairing interaction and the stacking interaction
\cite{PeyNLN}, i.e. those corresponding to our terms $V_p$ and
$V_s$. \EOR

\medskip\noindent
{\bf Remark 4.} Further support to this kind of approximation
comes from the results of \cite{CDG}, where the model with non
vanishing torsional interaction has been investigated in detail.
In that case, the torsional energy is a full order of magnitude
lesser then the stacking energy. Moreover, numerical soliton
solutions studied in that context showed that the form of the
soliton remains almost the same if $K_{t}=0$ and all the torsional
potential energy is transferred on the stacking term (see the Fig.
8 of \cite{CDG}).\footnote{At the kinematical level the only
difference between the model considered here and that in
\cite{CDG} is that here the base is a point particle whereas in
\cite {CDG} it is described by a disk of non vanishing radius.}
\EOR

\medskip\noindent
{\bf Remark 5.} We stress that, as quite common in DNA mechanical
models, we are dealing with the DNA molecule {\it per se}, i.e. we
do not consider interactions of the double chain with the solvent
(the fluid environment it is immersed in). This interaction would
lead to exchange of energy with the solvent, with dissipation and
random terms appearing in our dynamical equations (some authors
have taken into account interaction with the solvent by
introducing correction terms in the intrapair potential $V_p$
\cite{Web,ZC}). We hope a model modified in this direction can be
studied in the near future. \EOR

\subsection{Constraints and constraining potential}

It should be recalled that the angles $\th$ and $\phi$ are not on
the same footing. Indeed, while $\th$ is free to take any value,
$\th \in \R$, the angle $\phi$ is constrained to be in the range
\beq\label{constraint} \phi \ \in \ [- \phi_0 , \phi_0 ] \ . \eeq
The Euler-Lagrange equations (see below) for the Lagrangian
(\ref{lagr}) should be complemented with such a constraint.

Note that (\ref{constraint}) is an anholonomic constraint, and
taking it into account is somehow inconvenient. It may be
convenient, in particular in numerical computations, to remove the
constraint and introduce instead a {\it constraining potential}.
That is, we allow in principles $\phi$ to take any value, but
introduce a potential $U_c (\phi )$ which makes very costly
energetically any value of $\phi$ outside the range specified in
(\ref{constraint}), and has an irrelevant effect on the dynamics
when $\phi$ is within this range.

Any expression for the potential $U_c$ would be acceptable,
provided $U_c$ is essentially flat for $|\phi|< \phi_0$ and rises
sharply as soon as we approach the border -- or are outside -- of
this range. A convenient form for $U_c$ is \beq\label{conpot} U_c
(\phi) \ = \ \exp [ (K_c/2) (\phi^2 - \phi_0^2 )] \eeq with $K_c
>> 1 $. The total constraining potential appearing in the
Lagrangian will then of course be \beq V_c \ = \ \sum_a \sum_i U_c
( \phi^{(a)}_i ) \ . \eeq

\subsection{Equations of motion}

The equations of motion for the model are the Euler-Lagrange
equations for the Lagrangian (\ref{lagr}). These are obtained with
standard algebra, and are relatively involved expressions, which
we do not write explicitly. The equation for the dynamics of
variables associated with site $i$ involve also values of
variables at neighboring sites $i \pm 1$.

We are mainly interested in solutions which vary slowly in space
on the space scale set by the intersite distance $\de$. It is thus
convenient to pass to the continuum approximation; this means
promoting the array of values $\phi^a_i (t)$ to a field variable
$\Phi^a (z,t)$ such that $\Phi^a (n \de, t) = \phi^a_n (t)$, and
similarly promoting $\th^a_i (t)$ to a field variable $\Theta^a
(z,t)$ such that $\Theta^a (n \de , t) = \th^a_n (t)$. (Note we
are keeping the dimensional constant $\de$, so to have physical
units -- rather chain units -- for $z$.)

Moreover, in order to focus on the essential features of the model
-- and similarly to what is traditionally done in studying the
Peyrard-Bishop and Barbi-Cocco-Peyrard models
\cite{BCP,BCPR,PeyNLN,PB} -- we will just consider symmetric
motions; that is, we assume $ \phi^{(+)}_i (t) = \phi^{(-)}_i
(t)$, $\th^{(+)}_i (t) = \th^{(-)}_i (t) $ which of course entails
equality of the field variables as well. We will thus write simply
$ \Phi^{(\pm)} (z,t) = \Phi (z,t)$, $\Theta^{(\pm)} (z,t) = \Theta
(z,t)$; and similarly for derivatives.

\medskip\noindent
{\bf Remark 6.} It should be stressed that if we restrict to
symmetric solutions in discussing the standard Y model, then only
one type of kink solitons can exist, i.e. those connecting
equilibrium positions at $x = \pm \infty $ with a total phase
shift of $\pm 2 \pi$; higher solitons, even if with symmetric
topological numbers, would require a deviation from a fully
symmetric dynamics. \EOR
\medskip

In this setting, we write $ \phi_i  \simeq \Phi$ and $\th_i \simeq
\Theta $ for variables at site $i$; and for neighboring sites we
have, expanding in Taylor series up to order two in $\de$, \beq
 \phi_{i \pm 1} = \Phi \pm \de \Phi_z + (1/2)
\de^2 \Phi_{zz} \ \ , \ \ \  \th_{i \pm 1} = \Theta \pm \de
\Theta_z + (1/2) \de^2 \Theta_{zz} \ . \eeq

With these, the Euler-Lagrange equations issued from (\ref{lagr})
are rewritten as second order PDEs for $\Phi$ and $\Theta$. In
order to simplify the writing, we introduce the notations \beq \a
(\Phi) \ := r^2 + r R \cos ( \Phi ) \ ; \ \ \b (\Phi ) \ := \ r^2
+ R^2 + 2 r R \cos (\Phi ) \ . \eeq The Euler-Lagrange equations
are then given explicitly by \beq\label{eulerlagrange}
\begin{array}{l}
\de^2 K_s (r^2 + r R \cos \Phi) \, \Phi_{xx} \ + \ \de^2 K_s (r^2
+ R^2 + 2 r R \cos \Phi) \,
\Theta_{xx} \ + \\
\ \  - \ [ M R^2 + m (r^2 + R^2 + 2 r R \cos \Phi)] \,
\Theta_{tt} \ - \ m (r^2 + r R \cos \Phi) \, \Phi_{tt} \ = \\
\ \ \ = \ r R [ (\de^2 K_s \sin \Phi ) \Phi_x (\Phi_x + 2
\Theta_x) \ - \ (m \Phi_t (\Phi_t + 2 \Theta_t) ] \ + \\
\ \ \ + \ 4 K_p (r R + R^2 ) \sin \Theta \ + \
4 K_p (r R + r^2 ) \sin (\Phi + \Theta ) \ ; \\
\de^2 K_s r^2 \, \Phi_{xx} \ + \ \de^2 K_s (r^2 + r R \cos \Phi) \, \Theta_{xx} \ + \\
\ \  - \ m r^2 \, \Phi_{tt} \ - \ m (r^2 + r R \cos \Phi) \, \Theta_{tt} \ = \\
\ \ \ = \ - \ (r R \sin \Phi) (4 K_p - m \Theta_t^2 + \de^2 K_s \Theta_x^2) \ + \\
\ \ \ + \ 4 K_p r (r + R) \sin (\Phi + \Theta ) \ + \ K_c \, \exp
[(K_2/2) (\Phi^2 - \phi_0^2)] \, \Phi \end{array} \eeq

\section{Travelling wave solutions (solitons)}

We are specially interested in travelling wave solutions, i.e.
solutions satisfying \beq\label{tws} \Phi (z,t) \, = \, \vphi
(z-vt) \, \equiv \, \vphi (\xi) \ , \ \ \Theta (z,t) \, = \, \vth
(z-v t) \, \equiv \, \vth (\xi ) \ .  \eeq These will be called
``solitons'' -- provided they satisfy suitable boundary
conditions, see below -- following the common language in DNA
modelling, and are indexed by an integer (i.e. a topological
winding number). We stress that they are not proven to be {\it
dynamical} solitons in the rigorous mathematical sense
\cite{CalDeg}; on the other hand, they are {\it topological}
solitons in the sense of field theory \cite{DNF}.

\subsection{Equations of motion}

With this {\it ansatz}, the equations (\ref{eulerlagrange}) read
\beq\label{elsol} \begin{array}{l}
\mu (r^2 + r R  \cos \phi) \ \phi'' \ + \ (J + 2 \mu r R \cos \phi) \ \theta'' \ = \\
\ \ \ \mu (r R \sin \phi ) \, [(\phi ')^2  + 2 \phi' \theta'] \ -
\ 4 K_p [(r R + R^2) \sin \theta \, + \, (r R + r^2 ) \sin (\phi + \theta)] \ ; \\
\mu r^2  \, \phi'' \ + \ \mu
  (r^2 + r R  \cos \phi) \, \theta'' \ = \
  (4 K_p r R  + \mu r R (\theta')^2 ) \, \sin \phi \ + \\
  \ \ \ - \ 4 K_p r (r + R) \, \sin (\phi + \theta) \ + \ K_c \,
  \exp[(K_c/2) (\phi^2 - \phi_0^2)] \, \phi \ .
\end{array} \eeq
Here we have simplified the writing by introducing the parameters
\beq\label{parameters} \mu \ := \ m \, v^2 \ - \ K_s \, \de^2 \ ,
\ \ J \ = \ (\mu (r^2 + R^2) + M R^2 v^2) \ . \eeq In the analysis
of the (\ref{elsol}), we will proceed perturbatively; see next
section.

\subsection{Finite energy condition and boundary conditions}

The equations (\ref{elsol}) are a reduction of
(\ref{eulerlagrange}), which are in turn issued by the Lagrangian
(\ref{lagr}). The latter is well defined on fields $\Phi (z,t),
\Theta (z,t)$ satisfying a {\it finite energy condition}; in the
continuum approximation the action $S$ is given by integration
over $z \in (-\infty,+\infty)$ of the Lagrangian density obtained
as the continuum limit of (\ref{lagr}). In the present case, where
we have a kinetic and a potential energy parts, the finite energy
condition corresponds to requiring that for large $|z|$ the
kinetic energy vanishes and the configuration correspond to points
of minimum for the potential energy.

By the explicit expression of the kinetic energy and of our
potentials, this means \beq\label{finener}
\begin{array}{l} \Phi (\pm \infty,t) = 0 \ , \ \Theta (\pm \infty,t)
= 2 n_\pm \pi \ , \\
\Phi_z (\pm \infty,t) = 0 \ , \ \Theta_z (\pm \infty,t) = 0 \ , \\
\Phi_t (\pm \infty,t) = 0 \ , \ \Theta_t (\pm \infty,t) = 0 \ ;
\end{array} \eeq where of course $\Phi(\pm \infty,t)$ stands for
$\lim_{z \to \pm \infty} \Phi (z,t)$, and so on.

When we pass to the travelling waves equation, this means that we
must impose on the functions $\vphi (\xi)$, $\vth (\xi )$ the
boundary conditions \beq\label{fes} \vphi (\pm \infty) = 0 \ , \
\vth (\pm \infty) = 2 n_\pm \pi \ , \ \ \vphi' (\pm \infty) = 0 \
, \ \vth' (\pm \infty) = 0 \ . \eeq

In the following we will describe the dynamics of $\vth (\xi),
\vphi (\xi)$ in terms of motions (in the fictitious time $\xi$) in
an effective potential; note that such a motion can satisfy the
boundary conditions (\ref{fes}) only if $(\vphi,\vth) = (0,2 \pi k
)$ is a point of maximum for the effective potential. The
solutions satisfying (\ref{fes}) can hence be classified by the
winding number $n := n_+ - n_-$.

In the following we will in particular attempt to describe the
basic ($n=1$) soliton; note that -- as stressed in Remark 6 --
this is the only case for which we can fully compare the solutions
of our model with those of the standard Y model within the frame
of symmetric solutions.

\section{Perturbative expansion around \Y solitons}
\label{perturbative}

As remarked above, while ($\th$ and hence $\Theta$ and) $\vth$ can
take any value, the range of ($|\phi|$ and hence $|\Phi |$ and)
$|\vphi|$ is limited by $\phi_0$; thus if $\phi_0 = \eps << 1$, we
are guaranteed $\vphi$ will also be of order $\eps$.

On the other hand, for $\phi_0 \to 0$ and hence $\vphi \to 0$, the
$\phi$ degree of freedom is frozen and we are effectively back to
consider the Y model \cite{CDG}; we would then obtain the soliton
solutions to our model as a perturbation of the corresponding
soliton solutions for the Y model. The Y model can be also
recovered from our composite model by acting on the geometrical
parameter characterizing it \cite{CDG}; this essentially amounts
to a suitable limit $r=0$. Here ``suitable'' means that some care
should be taken in the rescaling in order to avoid a singular
limit arises.

We want to study the equations of motion for our model in the
regime of small $\eps$, by looking at them as a perturbation of
the standard \Y equations. We stress that we are considering the
symmetric (in the chain exchange) setting, and adopting the
contact approximation $A=R+r$.

We will take $\phi_0 = \eps <<1$ and expand in $\eps$. We will
also take $r = \eps r_0$ and, in order to keep the total length of
the double pendulum $A=R + r$ constant, $R = A - \eps r_0$.

\medskip\noindent
{\bf Remark 7.} As shown in \cite{GY1}, one can obtain analytic
results also without letting $\ell_0 \to 0$; on the other hand it
is also shown there that a nonzero $\ell_0$ does actually produce
a very small modification in the solitons' shape, but yields much
more involved analytic formulas. Thus, in the present context it
is convenient to adopt $\ell_0=0$. Note that in any case we should
keep the physical distance $A$ fixed as $\eps$ is varied. \EOR
\medskip

We will expand $\vphi$ and $\vth$ to order $\eps$. We
write\footnote{The reader might me puzzled, as the title of this
section seems in contradiction with the expansion
(\ref{expansions}): \Y soliton involves only the topological angle
$\vth$, and here we are not even assuming $\vth_0$ is the \Y
soliton. We will soon see that $\psi = 0$ and $\vth_0$ is
precisely the \Y soliton as a consequence of the equations of
motion, i.e. (\ref{expansions}) leads to an expansion around the
\Y soliton with no need for a specific assumption.}
\beq\label{expansions}
\begin{array}{l} \vphi = \psi + \eps \vphi_0 \
+ O(\eps^2 ) \ , \\
\vth = \vth_0 + \eps \eta \ + O(\eps^2 ) \ .
\end{array} \eeq

We plug this into equation (\ref{elsol}) and expand the equations
so obtained to first order in $\eps$, obtaining the equations
\beq\label{seriesprel}
\begin{array}{l}
A^2 ( (\mu - M v^2) {\vth_0}'' - 4 K_p \sin \vth_0) +
  A \[ - (\mu - M v^2) (2 r_0 {\vth_0}'' - A \eta'' ) + \right. \\
\ \ \ \left. +  r_0 (\psi'' + 2 \mu {\vth_0}'') \cos \psi -
        \psi' r_0 (\psi' + 2 {\vth_0}') \mu \sin \psi
        - 4 K_p r_0 \sin \vth_0 + \right. \\
\ \ \ \left. - 4 K_p (A \eta \cos \vth_0
        - 2 r_0 \sin \vth_0) - 4 K_p r_0 \sin (\psi + \vth_0) \]
        \eps \ = \ 0 ; \\
- \exp [(K_c/2) \psi^2 ] K_c \psi \ + \ \[ A \mu r_0 (\cos \psi)
{\vth_0}'' +
  A r_0 (4 K_p + \mu ({\vth_0}')^2) \sin \psi + \right. \\
  \ \ \ \left. - 4 A K_p r_0 \sin (\psi +
  \vth_0)
  - \exp [(K_c/2) \psi^2 ] K_c (1 + K_c \psi^2 ) \vphi_0 \] \eps \
  = \ 0 \ . \end{array} \eeq

At order zero the second equation in (\ref{seriesprel}) yields
simply \beq\label{psi} \psi \ = \ 0 \ . \eeq With this, the
equations (\ref{seriesprel}) become quite simpler, i.e. reduce to
\beq\label{series}
\begin{array}{l}
A^2 ( (\mu  - M v^2) {\vth_0}'' - 4 K_p \sin \vth_0) + \\
\ \ \ +  A \[ 2 M r_0 v^2 {\vth_0}'' + A (\mu - M v^2) \eta''  -
        4 A \eta K_p \cos \vth_0 \] \ \eps \ = 0 \ ; \\
\[ - K_c \vphi_0 + A \mu r_0 {\vth_0}'' - 4 A K_p r_0 \sin \vth_0
\] \ \eps \ = \ 0 \ .
\end{array} \eeq

We will now compute the leading term contributions for $\vth$ and
for $\vphi$; this will suffice to show our main point, emphasized
in section \ref{slaved}. See Appendix B for the $O(\eps )$
correction to $\vth$.

\subsection{The leading term for $\vth$}

As for the first of (\ref{series}), at order zero we get a
sine-Gordon equation, \beq\label{orderzero}
 {\vth_0}'' \ = \ - (4 K_p R^2 /J) \ \sin \vth_0 \ . \eeq
This can be seen as describing the motion (in the fictitious time
$\xi$) of a point particle of unit mass in the field of an
effective potential \beq\label{V0} V_0 \ = \ {4 K_p R^2 \over J} \
(1 - \cos \vth_0 ) \ , \eeq where the additive constant has been
chosen so that for $\vth_0 = 0$ we get $V_0 =0$.

It is important to stress that the leading term of the
perturbative expansion reproduces the Y soliton equation
(\ref{orderzero}) without any further constraint. In particular
there is no constraint on the travelling  speed of the soliton.

\bigskip\noindent
{\bf Remark 8.} This behavior should be compared with that
described in \cite{CDG} for a similar model, which differs from
that under consideration here, essentially for the presence of a
non vanishing torsional potential. In this latter case setting
$\vphi=0$ allows to recover the Y soliton but the speed of the
travelling wave is fixed in terms of the torsional constant
$K_{t}$; moreover, the solitonic solution exists only when
$K_{t}/I<K_{s}/m$ \cite{CDG}. The fact that the above constraints
disappear when $K_{t}=0$ sheds light on the physical meaning of
the result of \cite{CDG}: the constraint fixing the soliton speed
found in \cite{CDG} has a purely dynamical origin and is
completely independent from both the geometry and the doubling of
the degrees of freedom introduced in the composite model; its
origin has to be traced back to the interplay between torsional
and stacking potential energy which characterizes the composite Y
model analyzed in \cite{CDG}. \EOR
\medskip

As discussed above, the boundary conditions (\ref{fes}) require
that $\vth_0 = 0$ corresponds to a maximum for the effective
potential; it is apparent from (\ref{V0}) that this is the case if
and only if $J < 0 $, which we assume from now on. By the
expression of $J$, see (\ref{parameters}), this is the case
provided \beq\label{vmax} v^2 \ < \ K_s \de^2 \ {r^2 + R^2 \over
(M+m) R^2 + m r^2 }  \eeq (the limiting speed for the standard Y
model is recovered for $r \to 0$). It should be noted that for $v$
satisfying (\ref{vmax}), the parameter $\mu$ is also negative,
$\mu < 0$ (more precisely, $- K_s \de^2 < \mu < - K_s \de^2 (1 +
(m/M) (1 + (r/R)^2) )^{-1}$.)

\begin{figure}
\begin{tabular}{cc}
 \includegraphics[width=180pt]{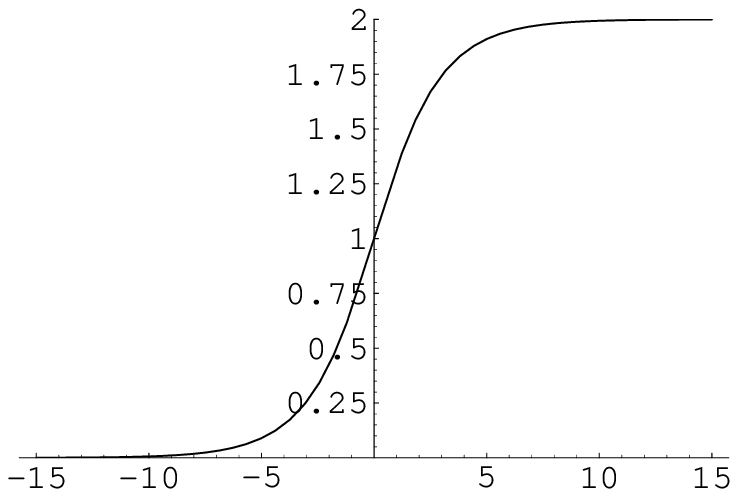} &
 \includegraphics[width=180pt]{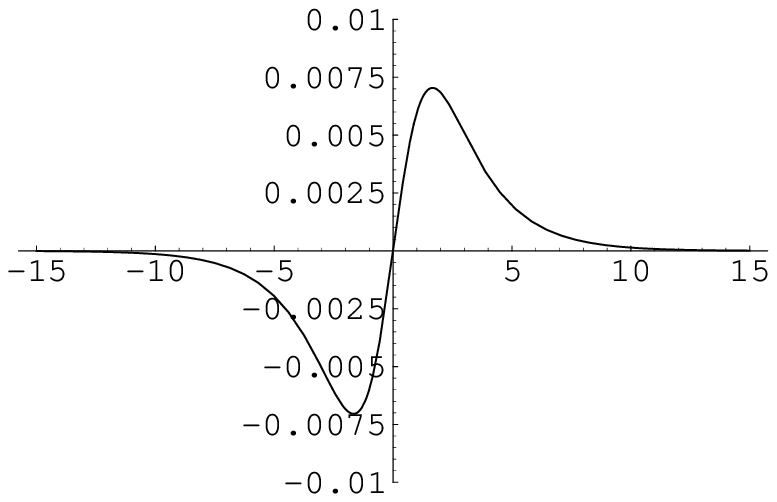} \\
 (a) & (b)
\end{tabular}
  \caption{Left (a): the function $\vth_0 (\xi)$, see (\ref{theta0}).
  Right (b): the function $\vphi_0 (\xi)$, see (\ref{phi02}). Here we adopted
the parameter values discussed in Appendix A, and plotted the case
$v=0$; we also used $K_c = 30$ in (b). The distance along DNA axis
$\xi$ is measured in units of the intersite distance $\de$, and
$\vth_0$, $\vphi_0$ in units of $\pi$.}\label{figt0p0}
\end{figure}

The ``conservation of energy'' in the potential $V_0$ for motions
satisfying (\ref{fes}) reads $(1/2) (\vth_0 ' )^2 + V_0 (\vth_0 )
= 0 $, i.e. we get $ {\vth_0}' = 4 R \sqrt{(K_p / |J|)} \, \sin
(\vth_0 / 2 )$. The equation is integrated explicitly by
separation of variables, yielding \beq\label{s0a}
\[ \log \( \tan (\vth_0/4) \) \] \ = \ K (\xi - \xi_0) \ , \eeq
where we have defined $ K := 2 R \sqrt{K_p/|J|}$; note that
choosing $\xi_0 = 0$ corresponds to requiring that $\vth_0 (0) =
\pi$. With this, and inverting (\ref{s0a}), we have the solution
(plotted in fig.\ref{figt0p0}.a) \beq\label{theta0} \vth_0 (\xi )
\ = \ 4 \ {\rm arctan} \[ \exp \( K \xi \) \] \ . \eeq

\subsection{The leading term for $\vphi$}

Let us now pass to consider the $O (\eps)$ terms in
(\ref{series}). In particular, the second equation reads
\beq\label{phi01} K_c \vphi_0 + 2 q R (4 K_p \sin \vth_0 + \mu
{\vth_0}'' ) \eeq and yields directly an expression for $\vphi_1$
in terms of $\vth_0$. Using the equation (\ref{orderzero}), this
is \beq\label{phi02} \vphi_0 \ = \ - {8 K_p q R (\mu R^2 - J)
\over J K_c } \ \sin \vth_0 \ . \eeq This function is plotted in
fig.\ref{figt0p0}.b.

\vfill\eject

\section{Slaved fields and realistic DNA modelling}
\label{slaved}

We have thus obtained the first order term of the $\eps$ expansion
for $\vphi$ in terms of the $\vth_0$ solution, with an algebraic
rather than differential equation, see (\ref{phi02}). In other
words, $\vphi_0$ is slaved to $\vth_0$; the same could be said
more generally for the field $\Phi$ with respect to the field
$\Theta$.

It is quite apparent that this feature is due to the different
nature of the fields in our model. As already pointed out,
$\Theta$ is free to change and go round the circle (is
topological), while $\Phi$ is constrained to small variations.
When we look for finite energy solutions, moreover, $\Theta$ will
join different vacua at $z = \pm \infty$, while $\Phi$ can at most
have small oscillations around the zero value.

If we look at this feature from the point of view of DNA
modelling, one point should be stressed. That is, suppose we had a
non-homogeneous model, the inhomogeneity being however such to
affect only the dynamical parameters related to the interaction
between $\phi_i$ angles. In this case, the constants appearing in
(\ref{phi02}) would not be constant any more, and actually depend
on the $z$ coordinate -- in the travelling wave reduction, on
$\xi$ and also on the parameter $v$ describing the speed of the
travelling wave. This would lead to a more involved expression for
$\vphi_0$ in terms of $\vth_0$, now depending explicitly on $\xi$
as well as on $v$, and with all parameters being a function of
$\xi$; but the relevant point is that we would still be able to
express $\vphi_0$ in terms of $\vth_0$.

Needless to say, this is only true at first order, and higher
order computations would soon become very much involved. On the
other hand, it remains true that the ``master'' dynamics would be
the one related to the $\theta$ angles, and the $\phi$ angles
would remain slaved at all orders; that is, writing
$$ \vth = \vth_0 + \eps \sum_{k=0}^\infty \eps^k \eta_k \ , \
\vphi= \sum_{k=0}^\infty \eps^k \psi_k \ , $$ the equation for
$\eta^k$ would depend on $\psi_0,...,\psi_{k-1}$ (beside depending
of course on $\vth_0$ and on $\eta_1,...,\eta_{k-1}$), while
$\psi_k$ would be determined algebraically by $(\vth_0 ,
\eta_1,...,\eta_{k-1} )$ in the $\vphi$ expansion.

The separation of the degrees of freedom in ``master'' and
``slaves'' is related to a nonperturbative feature of the
dynamical system (\ref{elsol}). The first integral  of Eqs.
\ref{elsol})  can be written in the form $\vphi'= F(\theta',
\theta,\vphi, E)$ where $E$ is an integration constant ( the
energy). Eliminating $\vphi''$ from the system (\ref{elsol}) and
using the first integral above, one can always write the resulting
equation in the form $G(\theta'',\theta', \theta,\vphi, E)=0$,
which defines in implicit form $\vphi$ as a  function of
$(\theta'',\theta', \theta,E)$. Thus, once a (perturbative or non
perturbative) solution for  $\theta$ is known, the corresponding
expression for $\vphi$ can be obtained algebraically.

\section{Discussion}
\label{discussion}

We have considered a mechanical model consisting of two chains of
double pendulums, coupled with each other at each chain site. This
``skeleton'' model is the simpler specimen of a class of models
for DNA roto-torsion dynamics which is aimed at improving the \Y
model, within the approach set forth by Englander et al.
\cite{Eng}, and should show the essential characteristics for the
class keeping to a minimum the unessential mathematical
complications.

An element of this class -- based on a geometrically more complex
and realistic description of the internal structure of DNA -- has
been considered in \cite{CDG}. We have introduced two
simplifications with respect to that model, which have been
suggested by numerical results for that model itself: bases have
been considered as point particles, and torsional interaction
disregarded. These two quite reasonable approximations simplify
drastically the dynamics of the system.

We have determined the PDEs describing the motions in the long
wavelength approximation, and also the reduction to travelling
wave solutions. The finite energy condition imposes boundary
conditions such that the finite energy solutions can be classified
by a topological integer -- actually a winding number for the
field $\Theta$ and hence for $\vth$ -- and minimal energy
solutions in each topological sector are topological solitons.

The equation of motion describing travelling waves for our
skeleton model are simple enough so that a (first order)
perturbative expansion around the well known \Y soliton can be
used to find approximate analytical solutions.

It seems appropriate to point out some drawbacks of our
discussion.
\medskip

\noindent (A) The question if our model equations support
dynamical solitons in rigorous mathematical sense \cite{CalDeg}
was not solved nor tackled.

\noindent (B) Similarly, we did not investigate the convergence of
the perturbation expansion around Y solitons, and just considered
first orders contributions.

\noindent (C) Also, as mentioned in Remark 5 and relevant for DNA
modelling, we did not consider interactions of the double chain
with the solvent.
\bigskip

On the other hand, we provided a simple model for a nonlinear
(double) chain showing some features which -- in our opinion --
are quite interesting both {\it per se} (i.e. in the frame of
nonlinear dynamics for discrete systems) and for applications, in
particular to DNA roto-torsional dynamics.

In particular, the model is simple enough to allow for an explicit
analysis, yet displays a very interesting interaction of
topological and non-topological degrees of freedom.

We have shown that the equations can be tackled in terms of a
perturbation expansion around the \Y soliton, and computed the
behavior of the new (non-topological) field at first order. Quite
remarkably, this is completely determined by the behavior of the
topological field at order zero. That is (in fluid dynamics
language), the field $\Phi$ appears to be {\it slaved} to the
field $\Theta$.

\section*{Acknowledgements}

This work received support by the Italian MIUR (Ministero
dell'Istruzione, Universit\`a e Ricerca) under the program
COFIN2004, as part of the PRIN project {\it ``Mathematical Models
for DNA Dynamics ($M^2 \times D^2$)''}.

\vfill\eject

\section*{Appendix A.  Parameters for DNA modelling}

Our model Lagrangian (\ref{lagr}) depends on several parameters.
Whatever the interest of the model {\it per se}, and albeit we
conducted our discussion in general terms, we are primarily
motivated by its application to describe the nonlinear dynamics of
DNA; hence we are interested in the values of these parameters for
the physical situation (also to be used in numerical plots later
on).

We will not discuss the derivation of these, referring to
\cite{CD,PeyNLN,Sae,Vol,YakuBook} as well as to the detailed
discussion given in \cite{CDG}.

As for the geometrical parameters, the physical values are
$$ A \simeq 10.5 \AA \ , \ R \simeq 6.0 \AA \ , \ r \simeq 3.0 \AA
\ . \eqno(A.1) $$ Note that $\ell_0 = 2 (A - R - r)$ is the length
of the pairing H-bonds at equilibrium; we have $\ell_0 \simeq 3.0
\AA$, in agreement with the physical situation \cite{Vol}.

In our discussion of the model, we have chosen to adopt
Yakushevich contact approximation $\ell_0 = 0$, which entails $A =
R + r$. We choose then to rescale $A$ by a factor (5/6), so to
satisfy this. Note that as our formulas only involve $r$ and $R$,
and not $A$, this does not show up in there.

The dynamic parameters are the effective masses $M$ and $m$; their
values are related with the moments of inertia $I_s$ and $I_b$ for
the rotations considered in the model. For these moments of
inertia we adopt the values \cite{CDG} $ I_s \simeq 2 \cdot
10^{-22} \, {\rm g \AA}^2$ and $I_b \simeq 5.5 \cdot 10^{-21} \,
{\rm g \AA}^2$, which yields for the effective masses the values
$$  M = I_s/R^2 = 5.5 \cdot 10^{-20} \, {\rm g} \ , \ \ m = I_b / r^2
\simeq 6.0 \cdot 10^{-20} \, {\rm g} \ . \eqno(A.2) $$

Finally, for the coupling constants we adopt (see again the
discussion in \cite{CDG}) the values
$$ K_p \ \simeq \ 0.022 {\rm eV/\AA}^2 \ , \ K_p r^2 \simeq 0.2
{\rm eV} \ , \ K_s \ \simeq \ 0.25 {\rm eV} \ . \eqno(A.3) $$

These values were used in the numerical computations and in the
plots of solution solutions given in the main text.

In particular, with these values of the physical parameters and
for $v=0$, the parameters $K = 2 R \sqrt{K_p / |J|}$ controlling
the soliton's shape, and $\a$ appearing in (B.1) below take the
value
$$ K \ \simeq \  0.53 \ ; \ \  \a \ \simeq \ 0.35 \ . \eqno(A.4) $$

\section*{Appendix B. Order $\eps$ corrections for $\vth$}

In the main text we provided the leading order terms for the
solution to (\ref{series}), i.e. a solution at order $\eps^0$ for
$\vth$, and at order $\eps$ for $\vphi$; it is appropriate to
mention that the $O(\eps)$ correction to the $\vth$ computed
above, call it $\eps \eta$, can be explicitly computed.

Indeed the $O (\eps )$ terms in the first of (\ref{series})
provides, using (\ref{theta0}), $$ \eta'' \ = \ - \ \a \, \cos
\vth_0 \ \eta \ - \ \b \, \sin \vth_0 \ ; \eqno(B.1) $$ we have
written, for ease of notation, $$ \a \, = \, {4 K_p / (\mu + M
v^2)} \ , \ \b \, = \, {8 K_p M r_0 v^2 / [A (\mu + M v^2)^2]} \ .
$$

We were not able to solve the equation (B.1) in general; however,
if we look at the case $v \to 0$, then $\b \to 0$ and the equation
reduces to $$ \eta'' \ = \ - \ {4 K_p \over \mu} \, \cos \vth_0 \
\eta \ . \eqno(B.2) $$

It is convenient to express $\eta (\xi) $ as a function $\rho
(\vth_0)$ of $\vth_0$ (which is a monotone function of $\xi$, as
seen above); with this, and with some simple algebra, the equation
(B.2) reads $$ 2 (1 - \cos \vth_0 ) \, \rho'' \ - \ ( \sin \vth_0
) \, \rho' \ + \ (\cos \vth_0 ) \, \rho \ = \ 0 \ . \eqno(B.3) $$
Note that we should as well require that $ \rho(0) = 0 = \rho (2
\pi )$, as the boundary conditions are satisfied by the first
order contribution $\vth_0$.

The solutions to (B.3) satisfying the boundary conditions are
given in terms of the hypergeometric function ${}_2 F_1$ by $$
\rho (\vth_0 ) \ = \ \sqrt{(1 - \cos \vth_0)} \ \  {}_2 F_1 \( {1
- i \sqrt{7} \over 4}, {1 + i \sqrt{7} \over 4}, 1, {1 \over 2} (1
- \cos \vth_0 ) \) \ ; \eqno(B.4) $$ this is readily converted
into $\eta (\xi)$ using (\ref{theta0}) and recalling that $$ \eta
(\xi ) \ := \ \rho [\vth_0 (\xi)] \ . \eqno(B.5) $$ Note it
follows from (\ref{theta0}) and trigonometric manipulations that
$$ \cos (\vth_0 ) \ = \ {1 - 6 x^2 + x^4 \over (1+x^2)^2} \ , \ \
{\rm where} \ x := \exp [K \xi ] \ . \eqno(B.6) $$

As for the stability of this solutions, it follows from our
selection of the function space to which solutions must belong,
i.e. from the boundary conditions at $\pm \infty$. In fact, we
note that as $\vth_0 (\xi) $ already satisfies the appropriate
limit conditions for $\xi \to \pm \infty$, the correction $\eta
(\xi ) $ should vanish for $\xi \to \pm \infty$; this implies the
stability of the solution.

The behavior of solutions for large $|\xi|$ can also be considered
as follows. For $\xi \to \pm \infty$ the term $\cos (\vth_0 )$
goes to 1, and asymptotically in $\xi$ eq.(B.2) reduces to
$$ \eta '' \ = \ {4 K_p \over |\mu|} \ \eta \ , $$
where we have used $\mu < 0$; needless to say, the solution of
this is given (in the $\xi \to \pm \infty$ region) by
$$ \eta \ = \ A_\pm \sinh (2 \sqrt{K_p/|\mu|} \xi ) \, + \, B_\pm
\cosh (2 \sqrt{K_p/|\mu|} \xi ) \ . $$ The limit conditions
require that $A_\pm = \pm B_\pm$.

\section*{Appendix C. Perturbative expansion for general
solutions}

We have considered expansion around Yakushevich soliton within the
function space ${\mathcal F}$ identified by the boundary
conditions (2.4) {\it and } by the travelling wave ansatz (2.1).
It is of course possible to consider expansion around the
Yakushevich soliton $\vth_0 (x-vt)$ also for functions which are
in ${\mathcal F}$ but are not necessarily travelling waves. In
this case the expansion (3.1) is replaced by (we could as well
insert a term $\Psi (x,t) $ of order one in $\Phi (x,t)$; the
vanishing of it would however obviously result from the equations
obtained at order $\eps^0$)
$$ \begin{array}{l}
\Phi (x,t) \ = \ \eps \, \phi (x,t) \, + \,
O(\eps^2) \ , \\
\Theta (x,t) \ = \ \vth_0 (x - v t) \, + \, \eps \, \eta (x,t) \,
+ \, O(\eps^2) \ . \end{array} \eqno(C.1) $$ Note that (2.4) imply
that the perturbation fields $\phi (x,t)$ and $\eta (x,t)$ must
vanish for $x \to \pm \infty$.

Inserting the (C.1) into the Euler-Lagrange equations (1.13) we
obtain at order $\eps$ two equations. One of these is actually an
algebraic relation between $\phi$ and $\vth_0$, which reads
explicitly
$$ \phi \ = \ - \ {4 A K_p M r_0 v^2 \over K_c ((M+m) v^2 - K_s \de^2 )}
\ \sin (\vth_0) \ . \eqno(C.2)$$ We stress that this imply that
the $\Phi$ field is slaved, even without the restriction to
travelling waves.

As for the other $O(\eps)$ equation, this reads
$$ (K_s \de^2) \, \eta_{xx} \, - \, (M+m) \, \eta_{tt} \ = \ 4 K_p
\, \cos (\vth_0 ) \, \eta \, + \, 8 {K_p M r_0 v^2 \over A((M+m)
v^2 - K_s \de^2 ) } \, \sin (\vth_0 ) \ . \eqno(C.3) $$ Here and
above we have used the writing $\sin (\vth_0 )$, $\cos (\vth_0)$
albeit these are known functions of $\xi = (x - v t)$, see (B.6),
for ease of writing.

For large $|t|$ and finite $x$ the (C.3) reduces to the autonomous
equation
$$ (K_s \de^2) \, \eta_{xx} \, - \, (M+m) \, \eta_{tt} \ = \ 4 K_p
\, \cos (\vth_0 ) \, \eta \ ; \eqno(C.4) $$ the dispersion
relations for this are easily obtained setting $\eta (x,t) = f_{k
\om} \exp [ i (q x - \om t ) ] $ and are given by
$$ \om \ = \ \pm \sqrt{{4 K_p + K_s \de^2 q^2 \over M+m}} \ . \eqno(C.5) $$


\label{lastpage}

\end{document}